# Anomalous exchange bias effect in ferromagnetic VI$_3$ flakes

Xi Zhang[1,2], Xiuquan Xia[2], Qiye Liu[4], Yonggang He[4], Le Wang[2,3], Junhao Lin[4], Jia-Wei Mei[4,6], Yingchun Cheng[5], Jun-Feng Dai[2,3,7*]

1. Frontiers Science Center for Flexible Electronics (FSCFE), Xi'an Institute of Flexible Electronics (IFE) and Xi'an Institute of Biomedical Materials & Engineering (IBME), Northwestern Polytechnical University, Xi'an 710072, China
2. Shenzhen Institute for Quantum Science and Engineering, Southern University of Science and Technology, Shenzhen, 518055, China
3. International Quantum Academy, Shenzhen, 518048, China
4. Department of Physics, Southern University of Science and Technology, Shenzhen, 518055, China
5. Key Laboratory of Flexible Electronics & Institute of Advanced Materials, Jiangsu National Synergetic Innovation Center for Advanced Materials, Nanjing Tech University, Nanjing 211816, China
6. Shenzhen Key Laboratory of Advanced Quantum Functional Materials and Devices, Southern University of Science and Technology, Shenzhen 518055, China
7. Shenzhen Key Laboratory of Quantum Science and Engineering, Shenzhen 518055, China

\* Corresponding authors:
daijf@sustech.edu.cn

**ABSTRACT**

**The exchange bias (EB) effect, pivotal in magnetic data storage and sensing devices, has been observed not only in interfacial regions but also in intrinsic ferromagnetic materials. Here, we've uncovered a robust and stable exchange bias effect within the layered van der Waals (vdW) ferromagnet VI$_3$ employing magnetic circular dichroism microscopy. At 10 K, we observed a significant exchange field of approximately 0.1 T, accompanied by random shifts (positive or negative relative to zero magnetic field) after zero-field cooling. Notably, this effect is effectively controllable after field cooling, with shift direction opposing the applied magnetic field. The presence of strong magnetic anisotropic energy within VI$_3$ results in larger coercivity-bound magnetic domains. These domains dictate the neighboring ferromagnetic alignment and induce shifts in the hysteresis loop. Our study not only contributes to comprehending fundamental nanoscale magnetic interactions but also sheds light on emergent phenomena within layered van der Waals magnets.**

Exchange bias (EB) effect was originally discovered in 1956 by Meiklejohn and Bean during their study of the magnetic properties of ferromagnetic Co particles within antiferromagnetic CoO material. This phenomenon is characterized by a shift in the hysteresis loop of the ferromagnetic (FM) state along the magnetic field axis, resulting from the interfacial exchange interaction between antiferromagnetic (AFM) and FM orders[1-3]. Its significance lies in its pivotal role in advancing potential applications within magnetic storage and logic devices. Over time, EB effect has been observed not only in the interface of ferromagnets and antiferromagnets but also in other magnetic combinations, such as FM/ferrimagnets[4] and FM/spin glass[5].

In recent years, research into van der Waals (vdW) magnets has experienced rapid growth, sparking interest in investigating EB within heterostructures composed of AFM and FM vdW magnets. The appeal of employing vdW magnets lies in their adjustable thickness, ideal for nanoscale spintronic devices, and their mechanical flexibility, suited for flexible electronic devices. Notably, experimental observations of EB have emerged in vdW $Fe_3GeTe_2/MnPX_3$[6], $Fe_3GeTe_2/CrCl_3$[7], and $Fe_3GeTe_2$/oxidized $Fe_3GeT_2$ heterostructures[8]. Furthermore, the EB effect has been documented in intrinsic vdW ferromagnets like $Fe_3GeTe_2$[9], where the coexistence of AFM and FM phases serves as a crucial factor. However, the microscopic mechanism behind this phenomenon in intrinsic vdW magnets remains a subject of ongoing debate. Contrary to the coupling observed among different magnetic orders, the pinning sites resulting from the defects[10, 11] or sit-mixing[12] also dominate the shift of hysteresis loop along the magnetic field, contributing significantly to a comparable EB effect.

In this study, we observed the EB effect in intrinsic FM $VI_3$ flakes utilizing reflected magnetic circular dichroism (RMCD) microscopy. Our observations revealed a random shift of the hysteresis loop in both positive and negative directions of the applied field following zero-field cooling, exhibiting an exchange field of up to ~0.1 T in thick $VI_3$ flakes at 10 K. Notably, we demonstrated the controllability of the shift direction through field cooling from paramagnetic to ferromagnetic states. The $V^{3+}$ ion in $VI_3$ exhibits a notably high magnetic anisotropy energy (MAE) due to its unquenched orbital moment and strong spin-orbit coupling[13], making it sensitive to the pressure[14, 15] and strain[16]. Consequently, the presence of local tensile strain tends to induce an increase in the local coercivity of domain, preventing reversal under low magnetic fields. These discrepant domains effectively pin the ferromagnetic order, thereby inducing a robust EB effect under an external field. Overall, our results contribute to deepening the understanding of the origin of the EB effect in intrinsic vdW ferromagnetism.

$VI_3$ belongs to the category of intrinsic layered ferromagnetic semiconductors with the Curie temperature of its single crystals estimated to be around 49.5 K[17], characterized by a canted FM order[18]. Within the $V^{3+}$ ion, the incomplete occupation of the $t_{2g}$ spin state induces an unquenched orbital moment, a consequence of its strong spin-orbit coupling. Consequently, the interplay between magnetic order and structure in $VI_3$ becomes notably intricate. Below Curie temperature, a distinctive FM state exhibiting a structure with lowered symmetry has been unveiled through specific heat and magnetization measurements[14], x-ray diffraction measurements[19], and neutron diffraction measurements[20]. Here, we employ the reflective magnetic circular dichroism (RMCD) to determine the magnetic properties of $VI_3$ flakes with an out-of-plane magnetic field. The measurements in this work were carried out under the excitation of a 633 nm HeNe laser. The laser power is limited to 10 μW to avoid any thermal effect. In addition, the RMCD mapping signal is collected by moving the laser point step-by-step on the sample, with a resolution of 1 μm under a fixed out-of-plane magnetic field.

We first characterize the magnetic spatial distribution of thin $VI_3$ flakes at the initial state using RMCD mapping after cooling the sample from the paramagnetic state without any applied field. Figure 1a shows two distinct types of magnetic domains with opposite magnetization orientations (red and blue regions) observed at 10 K, with the domain sizes of several micrometers. Remarkably, subsequent RMCD mappings under identical experimental conditions reveal a consistent magnetization pattern (Figure S1), indicating the negligible impact of laser heating on the observed magnetic orders. Upon the application of a uniform magnetic field in the +z direction, a predictable behavior emerges: the expansion of positive domains and contraction of negative domains in response to increasing field strength (Figure 1b-d). Notably, at approximate +0.2 T, a sharp increase in magnetization occurred (Figure 1e), resulting in the eventual disappearance of the multidomain structure at +0.3 T (Figure 1g). By analyzing the RMCD curve as a function of external magnetic field (Figure 1i), we can get the magnetization switching field in the range of 0.2 to 0.34 T in several thick samples. This is consistent with the depinning field of 0.1-0.25 T for $VI_3$ thick flakes revealed by nitrogen-vacancy (NV) microscope[21]. Importantly, upon removal of external magnetic field, the magnetic state can be nicely preserved with a permanent magnetization, as shown in Figure 1h. It suggests that the magnetocrystalline energy[13] is sufficiently larger than demagnetization field and Zeeman energy, contributing to stabilize and sustain the state of magnetization after the external magnetizing field is removed.

$VI_3$, known as a nucleation-type magnet[21], exhibits a sharp magnetization reversal as the external field approaches the coercive field. To explore the properties of nucleation centers, we conduct the RMCD measurements as a function of the external magnetic field under different magnetic field sweep ranges, which vary from ±0.5 T to ±2.5 T in a 0.5 T increments. Initially, applying a positive magnetic field of 0.5 T remove the multi-domain structure since it is greater than depinning field (0.3 T). The magnetic field sweep follows a cycle, starting from the zero field, increasing along positive direction to the maximum, decreasing to the maximum in the negative direction, and returning to zero. In Figure 2a, a rectangular hysteresis loop with sharp magnetization reversals (curves labelled as ±0.5 T 1$^{st}$ and 2$^{nd}$ in Figure 2a) is observed, where the coercivity is extracted to be around 0.40±0.1 T at 10 K. This transition suggests that the sample has transformed from the multi-domain structures to single phase uniaxial magnet, behaving as a large single domain particle with strong out-of-plane magnetic anisotropy. When extending the sweep range to ±1 T and following the same cycle, the original magnetization jump at positive direction persists around 0.44 T. Significantly, the reversal at negative field direction increases from initial -0.4 T to -0.84 T with a near doubling of the coercivity (curve labelled as ±1 T 1$^{st}$ in Figure 2a). Upon sweeping in the range of ±1 T again, the original magnetization jump at 0.44 T vanishes, replaced by a new coercivity with a great magnitude of 0.73 T (curve labelled as ±1 T 2$^{nd}$ in Figure 2a), while the one at negative direction is around -0.87 T. Similar phenomena occur with ±1.5 T sweep range (curves labelled as ±1.5 T 1$^{st}$ and 2$^{nd}$ in Figure 2a), showing an ultimate coercivity at around 0.97 and -1.12 T, receptively. Further increasing the ranges to ±2 and ±2.5 T (Figure 2b), the coercivity does not change much and eventually stabilize. Therefore, increasing the sweep range or the magnetizing field, induces a rise in coercivity from ~0.40 at 0.5 T to ~1 T at 1.5 T. This could be due to the presence of residual reverse domains acting as nucleation centers that resist magnetization at low fields, thus aiding magnetization reversal and reducing coercivity. As the applied field strengthens, these residual domains vanish, leading to the disappearance of magnetic reversal at low field[22]. It's worth noting a similar phenomenon in ref. [[21]], where low fields fail to properly magnetize 3-layer $VI_3$ flakes revealed by nitrogen-vacancy (NV) spectroscopy. In addition, it is worth noting that the presence of a hysteresis loop at low fields is not universally observed across all samples. As depicted in Figure S2, among the five $VI_3$ flakes examined, two samples exhibit a hysteresis loop during low-field sweeping (within ±0.5 T). Moreover, after elevating the temperature above $T_c$ and subsequently cooling the sample to 10 K in a zero magnetic field, the hysteresis loop remains unobservable within a sweeping field of ±0.5 T (Data not shown). These observations further support our explanation regarding the presence of unstable reverse domains.

It's intriguing that the hysteresis loop exhibits a negative shift relative to zero magnetic field across sweep ranges of ±1.5, ±2, and ±2.5 T, as depicted by the symmetrical dashed rectangles in Figure 2b. To clarify its origin, our focus shifts to examining the phenomenon of hysteresis loop shifts after zero-field cooling. Prior to each measurement, a 1.5 T positive magnetic field is applied to the $VI_3$ flakes, aimed at minimizing multidomain and weak coupling states. In Figure 3a, a noticeable shift of the hysteresis loop towards the negative direction becomes apparent, leading to an asymmetrical pattern. To quantify this effect, we define exchange-bias field ($H_{ex}$) as ($\frac{H_{cl}+H_{cr}}{2}$), where $H_{cl}$ ($H_{cr}$) represents the coercivity at negative (positive) magnetic fields. Over six consecutive measurements, the negative shift can remain with a fluctuation. This behavior sharply differs from other 2D magnets like $Fe_3GeTe_2$, where a training effect leads to the diminishing or disappearance of the shift in consecutive hysteresis loop measurements[23]. The extracted $H_{ex}$ from the asymmetrical hysteresis loop in six successive measurements is approximately 0.12 ± 0.03 T, roughly twice the value observed in $Fe_3GeTe_2$[7,8]. In addition, despite attempts to vary sweeping directions and initial magnetization, the shift direction of the hysteresis loop after zero-field cooling remains unaffected (Figure S3). The apparent explanation lies in the existence of interlayer AFM order, which anchors the adjacent FM structures. However, our field cooling magnetization measurement on $VI_3$ crystal unequivocally exhibit ferromagnetism without any observable AFM orders (Figure S4). Furthermore, the DFT calculations in ref. [24] affirm robust ferromagnetic ground state in VI3, irrespective of stacking configurations. Additionally, ref. [25] reports interlayer coupling favoring antiparallel spin states in another AB' stacking bilayer $VI_3$ with C2/m space-group symmetry. However, our polarization-resolved second harmonic generation (SHG) measurements illustrate a symmetrical six-petal pattern (Figure S5), indicative of $R\bar{3}$ symmetry. It indicates the absence of AB' stacking order within adjacent layers of $VI_3$ flakes. Hence, we can confidently exclude the influence of interlayer AFM coupling.

$VI_3$ is a ferromagnet with a relatively large MAE (~8 meV[13]) due to its strong spin-orbit coupling. As reported in ref. [16], the increased MAE under strain induces the enhancement of coercivity. Hence, if local magnetic domain arises due to localized tensile strain (as indicated by the blue arrow in Figure 3b), it tends to exhibit higher coercivity compared with other intrinsic regions. Consequently, it resists magnetization within a ±1.5 T external magnetic field. Figure 3b illustrates the micro-mechanism for the shift of hysteresis loop. The domain serves to fix neighboring magnetic orders in accordance with its specific orientation. When a opposite magnetic field is applied, it increases the magnitude of field required for the main phase magnetization, subsequently shifting the hysteresis loop. This behavior mirrors the exchange bias effect observed between FM and another magnetic states. An analogous exchange bias effect is also observed in a suspended $VI_3$ flake (Figure S6), signaling that the magnetic domain is unrelated to any influence from the substrate. The local magnetic domains are presumed to originate within the interior of $VI_3$ flakes. In contrast to instability observed in low-field hysteresis loops, the asymmetrical hysteresis loop around the zero field constantly appears in all tested $VI_3$ flakes, exhibiting relatively stable properties.

Surprisingly, upon successively heating the sample above its $T_c$ and subsequently cooling it at zero external field six times, the shift direction appears random (Figure 3c), resulting in either a positive or negative shift of the hysteresis loop. This randomness signifies that the orientation of magnetization of local domains is arbitrary after zero-field cooling. The scenario significantly differs after field cooling. Experimentally, the $VI_3$ flakes undergo a field-cooling process from room temperature to 10 K under an out-of-plane external magnetic field of ±1.5 T (either -1.5 T or +1.5 T), aligning the magnetization of FM state along the applied magnetic field direction. As depicted in Figure 4a, the hysteresis loop exhibits a distinct positive shift after negative field cooling and vice versa. Remarkably, the shift direction opposes the cooling field direction, yielding $H_{ex}$ values of approximately 0.26 T and -0.18 T for NFC and PFC (negative field cooling and positive field cooling), respectively. With increasing temperature, the extracted $H_{ex}$ value gradually decreases, disappearing around 40 K (denoted as $T_{ex}$) for both NFC and PFC cases. This temperature is lower than the disappearance temperature of the hysteresis loop, observed at 50 K (Figure 4a). This suggests that the development of local domains occurs after the formation of FM order (FM I) in VI$_3$ flakes. Around 40 K, there is another FM order (FM II) develops in VI$_3$, accompanying a structural transition from monoclinic to triclinic structure through x-ray powder diffraction[19]. Additionally, neutron powder diffraction reveals a further distortion of the honeycomb framework at around 32 K, shifting from a V-V dimer to three V-V bonds with distinct distances[19, 26, 27]. These observations suggest that local tensile strains tend to form in the transition from FM I to FM II states, generating an FM state with high coercivity.

**Conclusion**

Our investigation into the coercivity of VI$_3$ across various sweeping magnetic fields revealed distinct behaviors. Upon zero-field cooling, the material exhibited magnetic domains with sizes in the range of several micrometers, magnetizable beyond 0.3 external magnetic field strength. With an increase in the sweeping field from ±0.5 to 0.15 T, we observed a rise in coercivity. This points toward the existence of unstable reversal nucleation centers that lower the coercivity of the primary phase. Although the coercivity stabilizes above 1.5 T external magnetic field, an asymmetrical hysteresis loop emerges. Following zero-field cooling, the direction of shift in the hysteresis loop appears random, while during field cooling, it exhibits an opposite direction relative to the applied field, indicative of an exchange bias effect akin to that observed in the FM and AFM interface. We propose that domains characterized by larger coercivity, likely due to local tensile strain, contribute as magnetic centers. These centers pin the adjacent ferromagnetic order along their direction, resulting in a relatively large opposing field required to polarized the magnetization in VI$_3$ flakes.

**Author contributions**

J.D., Y.C., and J.M. conceived the project. X.Z., X.X., Q.L., Y.H. and J.D. designed and performed the experiments. L.W., and J.L. provided and characterized the samples. Y.C. and J.M. provided theoretical support. All authors discussed the results and co-wrote the paper.

**Acknowledgment**

We would like to thank Prof. Haizhou Lu from SUSTech for the helpful discussions. J.F.D. acknowledges the support from the National Natural Science Foundation of China (11974159) and the Guangdong Natural Science Foundation (2021A1515012316).

**Conflict of Interest**

The authors declare no conflict of interest.

**FIGURES AND CAPTIONS**

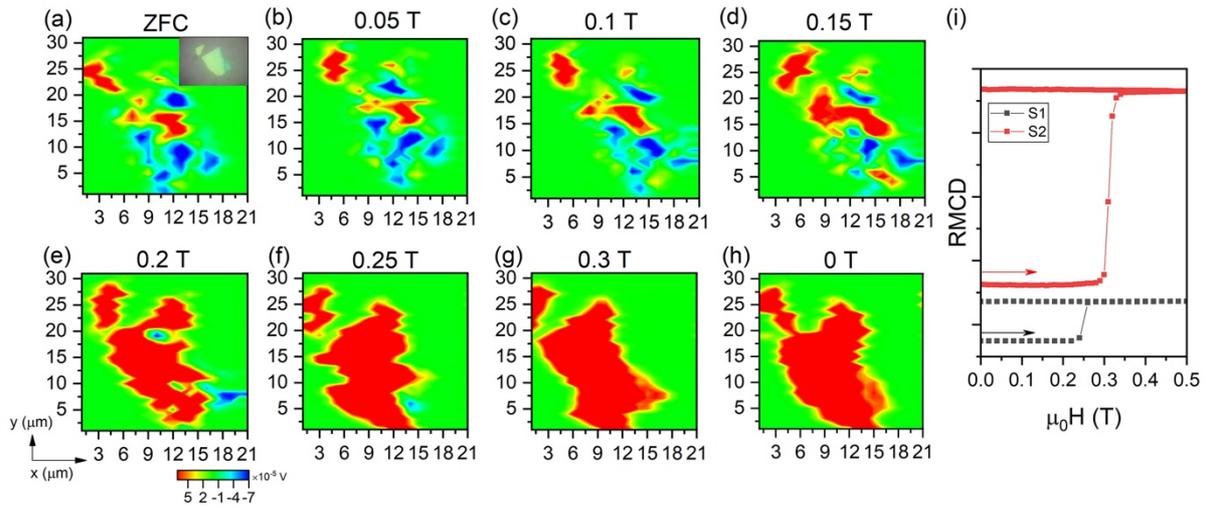

Figure 1: RMCD mapping at 10 K after (a) zero external field cooling and captured at various external magnetic fields of (b) 0.05, (c) 0.1, (d) 0.15, (e) 0.2, (f) 0.25, (g) 0.3, and (h) 0 T along the +z direction. The inset in (a) displays an optical image of a VI3 flake. (i) RMCD signal as a function of applied magnetic field in two VI$_3$ flakes, starting from zero field, as indicated by the arrows.

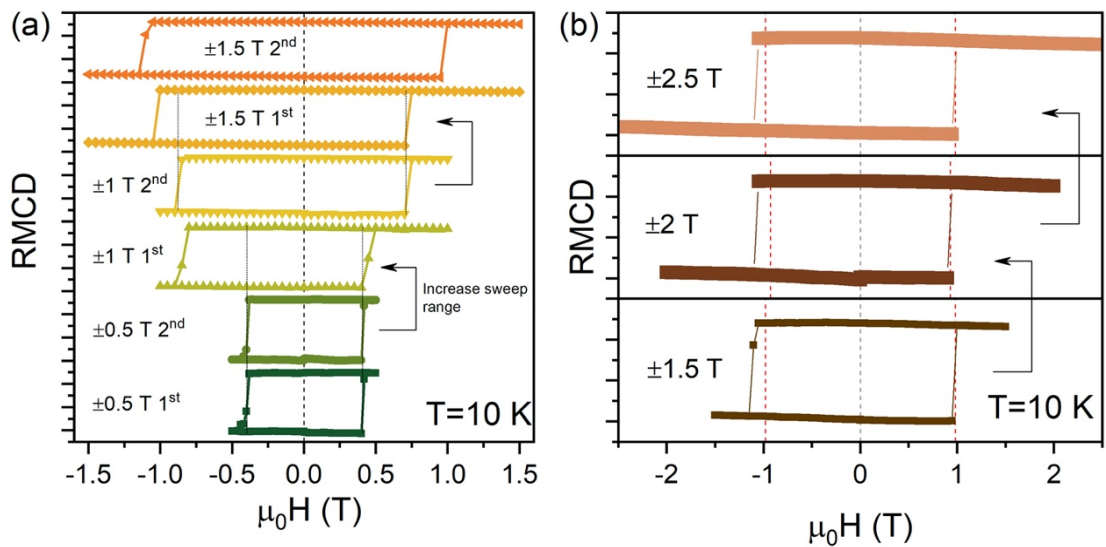

Figure 2: RMCD signal as a function of applied magnetic field at 10 K in a VI$_3$ flake across different sweeping ranges, e.g., (a) $\pm 0.5$ T and $\pm 1$ T, and (b) $\pm 1.5$ T, $\pm 2$ T, and $\pm 2.5$ T. The RMCD signals in (a) are measured twice consecutively. The arrows indicate to expand the sweep range.

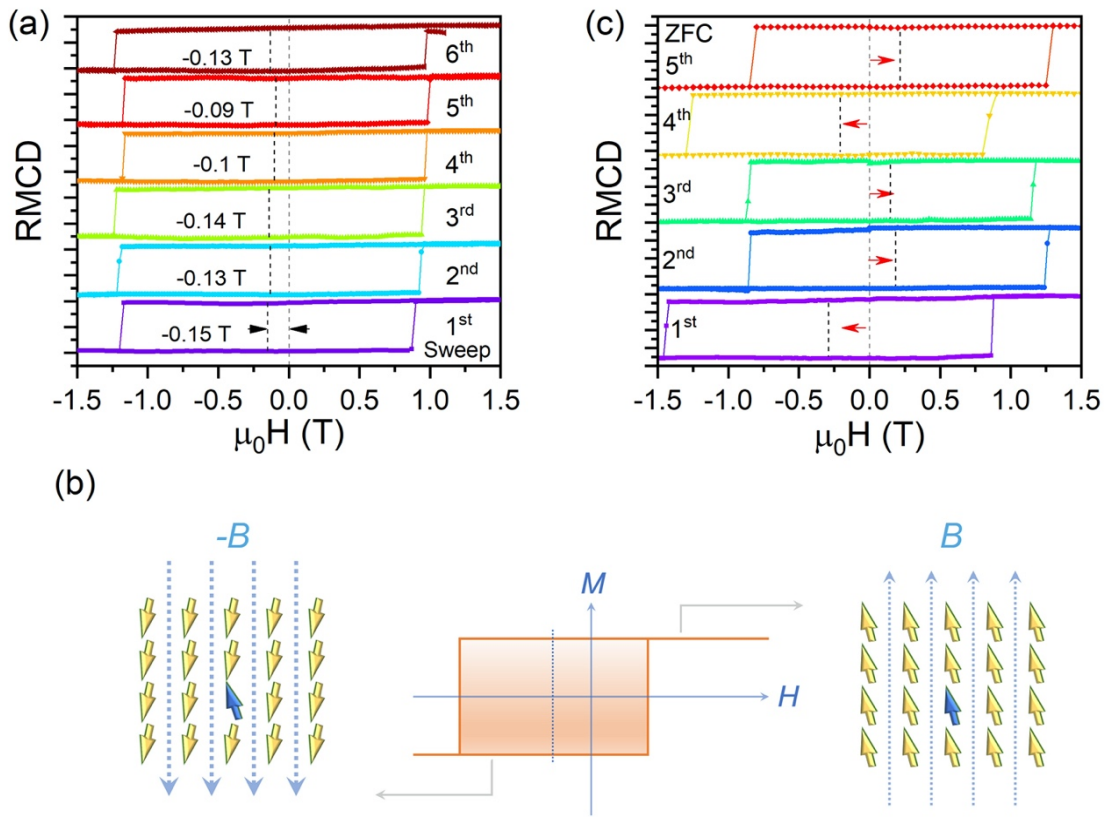

Figure 3: (a) RMCD signal as a function of applied magnetic field in consecutive six measurements. Prior to the measurement, the sample is magnetized at +1.5 T external field. (b) A schematic illustrating the mechanism behind the shift of hysteresis loop. (c) RMCD measurements conducted after five cycles of cooling from the paramagnetic to ferromagnetic states. The red arrows indicate the direction of shift relative to the zero field in the hysteresis loop.

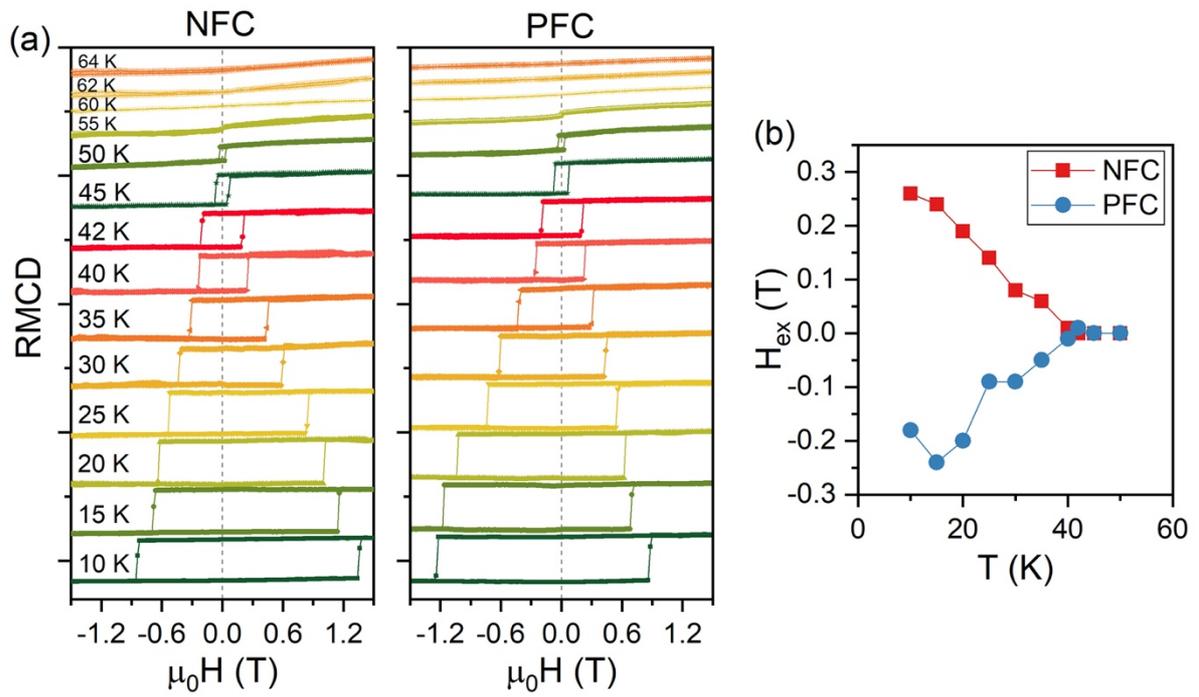

Figure 4: (a) RMCD signal as a function of applied magnetic field following positive (+1.5 T) and negative (-1.5 T) field cooling, respectively. (b) Temperature-dependent exchange bias field observed at NFC (red squares) and PFC (blue squares), respectively.

# Anomalous exchange bias effect in ferromagnetic VI$_3$ flakes

Xi Zhang[1,2], Xiuquan Xia[2], Qiye Liu[4], Yonggang He[4], Le Wang[2,3], Junhao Lin[4], Jia-Wei Mei[4,6], Yingchun Cheng[5], Jun-Feng Dai[2,3,7*]

1. Frontiers Science Center for Flexible Electronics (FSCFE), Xi'an Institute of Flexible Electronics (IFE) and Xi'an Institute of Biomedical Materials & Engineering (IBME), Northwestern Polytechnical University, Xi'an 710072, China
2. Shenzhen Institute for Quantum Science and Engineering, Southern University of Science and Technology, Shenzhen, 518055, China
3. International Quantum Academy, Shenzhen, 518048, China
4. Department of Physics, Southern University of Science and Technology, Shenzhen, 518055, China
5. Key Laboratory of Flexible Electronics & Institute of Advanced Materials, Jiangsu National Synergetic Innovation Center for Advanced Materials, Nanjing Tech University, Nanjing 211816, China
6. Shenzhen Key Laboratory of Advanced Quantum Functional Materials and Devices, Southern University of Science and Technology, Shenzhen 518055, China
7. Shenzhen Key Laboratory of Quantum Science and Engineering, Shenzhen 518055, China

## Supporting Information

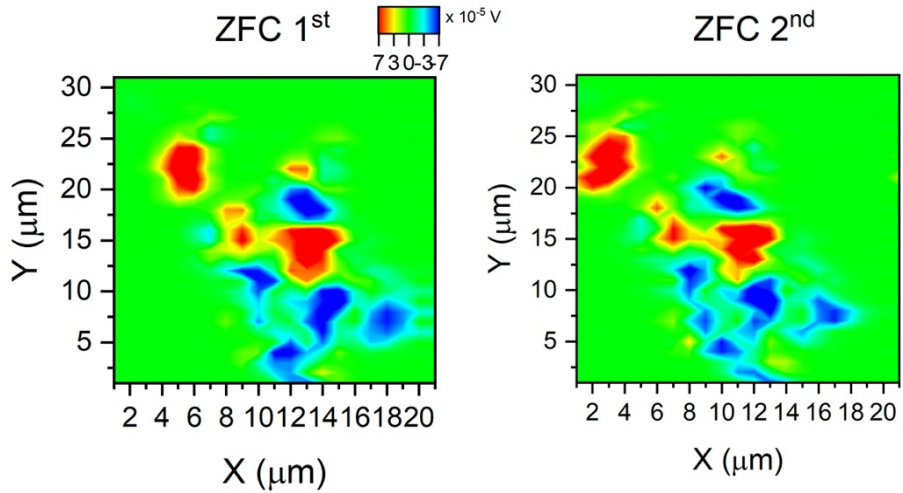

Figure S1: RMCD mapping conducted after zero field cooling in two consecutive measurements. Both mappings display similar domain pattern.

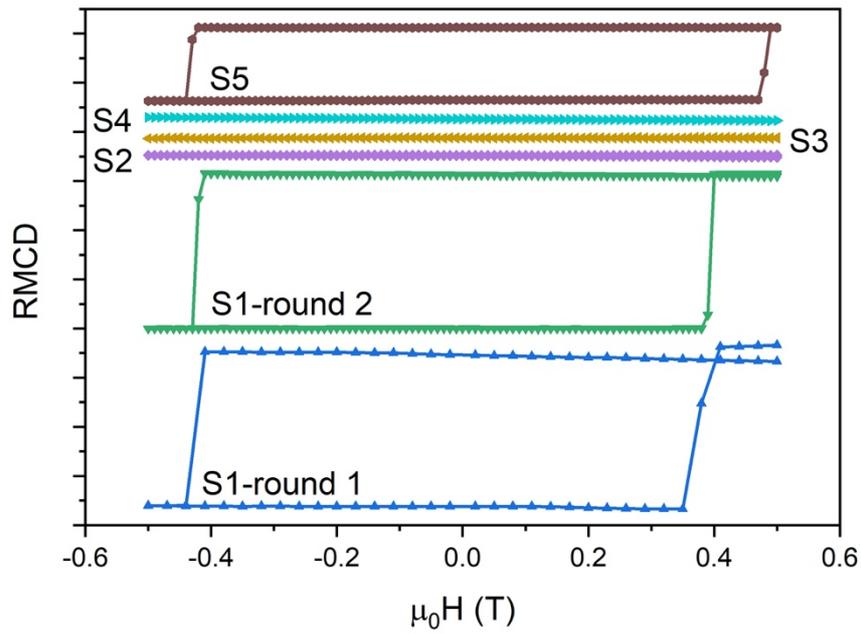

Figure S2: RMCD signal as a function of applied magnetic field within a sweeping range of ±0.5 T across five samples. Among these, two of the five samples exhibit hysteresis loops at a low sweeping range.

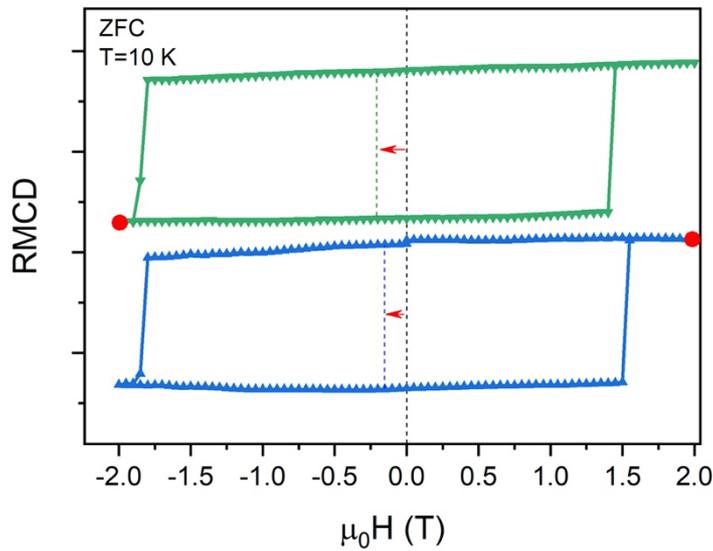

Figure S3: RMCD as a function of the applied magnetic field in $VI_3$ at 10 K. The blue curve represents the sweeping of the magnetic field starting from 2 T, while the green curve represents the sweeping starting from -2 T. The red points stand for the initial magnetic field for the sweep.

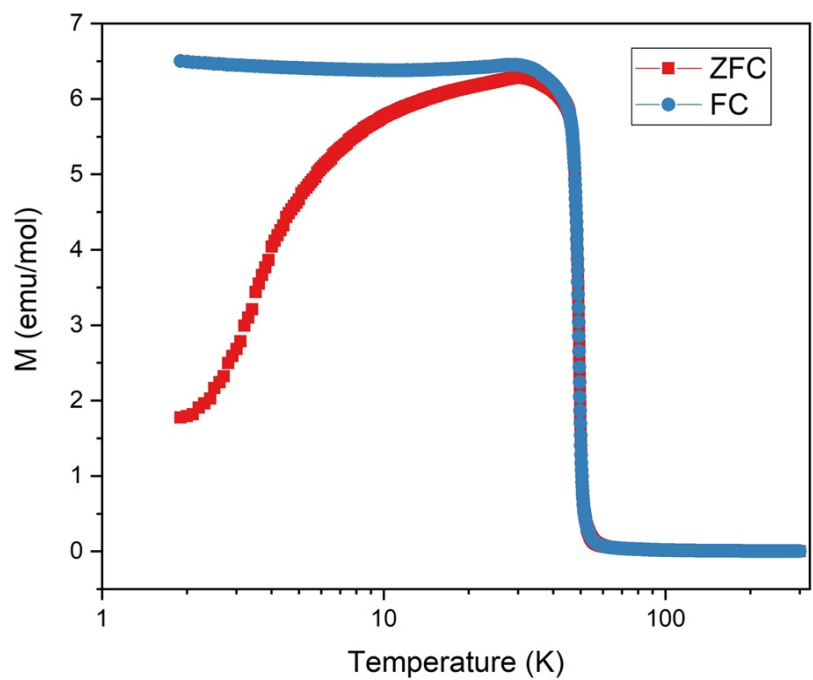

Figure S4: Temperature dependence of the zero-field cooling (ZFC) and field cooling (FC) magnetic susceptibility with a magnetic field of 0.05 T perpendicular to the ab-plane.

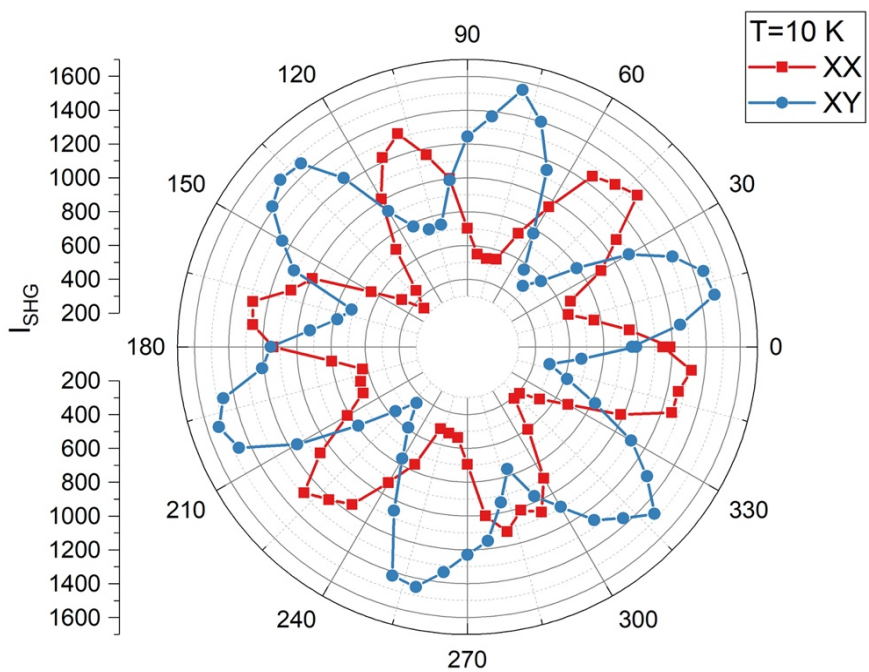

Figure S5: Polarization-resolved second harmonic generation (SHG) curve at 10 K, showcasing the parallel (XX) and perpendicular (XY) orientations between incident light and collected SHG signal.

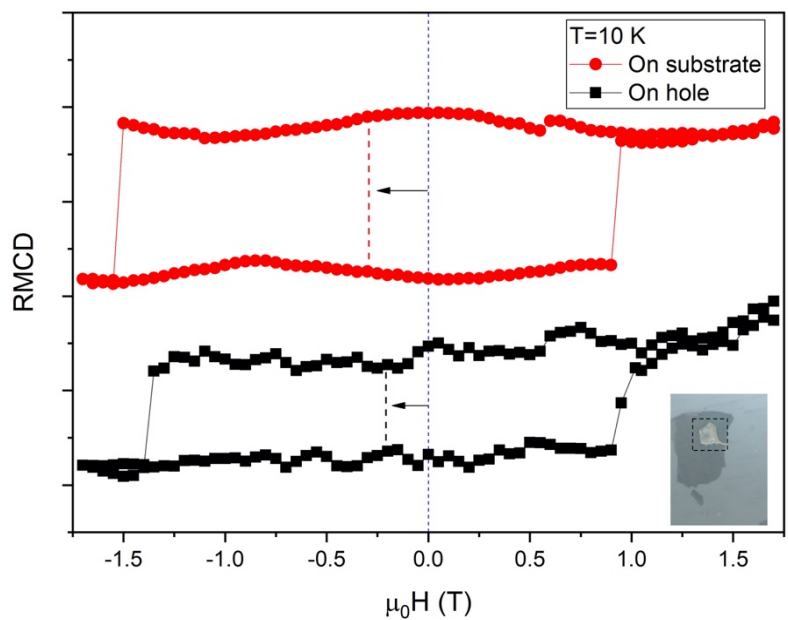

Figure S6: RMCD as a function of applied magnetic field for a suspended sample (black curve) and one placed on a substrate (red curve).